\documentclass[a4paper]{article}
\usepackage[colorlinks,linkcolor=red]{hyperref}
\usepackage{INTERSPEECH2021}
\usepackage{subfigure}
\newcommand{\tabincell}[2]{\begin{tabular}{@{}#1@{}}#2\end{tabular}}
\title{DCCRN+: Channel-wise Subband DCCRN with SNR Estimation for Speech Enhancement}
\name{Shubo Lv$^{*}$\thanks{*: Equal contribution. Lei Xie is the corresponding author.}, Yanxin Hu$^{*}$, Shimin Zhang, Lei Xie}
\address{
  Audio, Speech and Language Processing Group (ASLP@NPU), School of Computer Science, Northwestern Polytechnical University, Xi\textquotesingle an, China}
\email{\{shblv, yxhu, shmzhang\}@npu-aslp.org, lxie@nwpu.edu.cn}

\begin{document}

\maketitle
\begin{abstract}
    Deep complex convolution recurrent network (DCCRN), which extends CRN with complex structure, has achieved superior performance in MOS evaluation in Interspeech 2020 deep noise suppression challenge (DNS2020). This paper further extends DCCRN with the following significant revisions. We first extend the model to sub-band processing where the bands are split and merged by learnable neural network filters instead of engineered FIR filters, leading to a faster noise suppressor trained in an end-to-end manner. Then the LSTM is further substituted with a complex TF-LSTM to better model temporal dependencies along both time and frequency axes. Moreover, instead of simply concatenating the output of each encoder layer to the input of the corresponding decoder layer, we use convolution blocks to first aggregate essential information from the encoder output before feeding it to the decoder layers. We specifically formulate the decoder with an extra \textit{a priori} SNR estimation module to maintain good speech quality while removing noise. Finally a post-processing module is adopted to further suppress the unnatural residual noise. The new model, named DCCRN+, has surpassed the original DCCRN as well as several competitive models in terms of PESQ and DNSMOS, and has achieved superior performance in the new Interspeech 2021 DNS challenge. 
    
\end{abstract}

\section{Introduction}
    Speech enhancement refers to the task of eliminating background noise and improving speech quality and intelligibility from noisy audio signals. A decent speech enhancer has significant applications aiming to improve human or machine interpretation of speech, including hearing aids, audio communication and automatic speech recognition. Traditional speech enhancement methods usually apply a spectral suppression gain (or filter) to the noisy signal under the statistical signal processing theory. With the help of deep learning (DL), speech enhancement has been formulated as a supervised learning problem. Such data-driven approaches have become the mainstream because of their strong noise reduction abilities (especially for non-stationary noise) learned from simulated clean-noisy speech pairs. The recent deep noise suppression challenge (DNS) series\cite{reddy2020interspeech, reddy2021interspeech} have benchmarked many state-of-the-art DL-based speech enhancers, especially for real-time ones for speech communications, through subjective listening test and promising performance has been reported.
    
   A key challenge in developing a speech enhancer for human speech communication (and even for machine speech recognition) is how to preserve perceived subjective speech quality to the best extent while greatly suppressing the noise interference. Some recent studies, including the new DNS challenge initiative~\cite{reddy2021interspeech}, have pointed out that many neural noise suppressors are very good at suppressing noise, but do not improve the quality of speech, even introducing apparent speech distortions. In this paper, we study the problem by revising our recently proposed deep complex convolution recurrent network (DCCRN)~\cite{hu2020dccrn}.

    
    
      Mapping and masking are two commonly used DL-based strategies for speech enhancement. The mask-based approaches have gradually become the mainstream as they constraint dynamic range and usually converge faster. Various masks have been explored, including ideal binary mask (IBM)~\cite{wang2005ideal}, ideal ratio mask (IRM)~\cite{narayanan2013ideal} and spectral magnitude mask (SMM)~\cite{wang2014training}. Most of these approaches ignore the phase information as it is difficult to model due to its unclear structure. But recent speech enhancement studies have shown the clear benefits of modeling phase through phase-sensitive mask (PSM)~\cite{erdogan2015phase} and  complex ratio mask (CRM)~\cite{williamson2015complex}.
    In the CRM-based approach, the mask is applied to both real and imaginary (RI) components, and both magnitude and phase can be reasonably estimated, leading to improved enhancement performance. Time-domain approaches have also shown impressive noise suppression ability, which bypass explicit phase modeling by direct waveform input and output~\cite{luo2019conv}.

    Meanwhile, the speech enhancement performance has been boosted with the well-designed network structures~\cite{wang2015deep, pascual2017segan}. Recently, these network structures have gradually taken into account the phase information in the Time-Frequency (T-F) domain~\cite{choi2018phase}. Convolution recurrent network (CRN)~\cite{tan2018convolutional} is one of the most popular convolution encoder-decoder (CED) structures for speech enhancement. Originally it took a real spectrum as input and only estimated real mask through network. Later Tan \textit{et al.} proposed an updated structure of one encoder and two decoders for complex spectrum mapping (CSM)~\cite{tan2019complex} to estimate the real and imaginary parts of the spectrum, leading to improved performance. Very recently, we further updated the CRN network by introducing complex convolution and LSTM, resulting in the deep complex convolution recurrent network (DCCRN)~\cite{hu2020dccrn}. The DCCRN models are very competitive over other networks, either on objective or subjective metric. With 3.7M parameters, the model has ranked first in MOS evaluation in the Interspeech2020 DNS challenge real-time-track.

    In this paper, we further update DCCRN with the following important aspects, resulting in DCCRN+, aiming to improve perceived speech quality while greatly reducing the noise interference. 1) We enable the model with subband processing ability by learnable neural network filters. With smaller model size and speed-up inference, this update leads to 0.17 PESQ improvement as compared with the subband counterpart based on engineered FIR-filters, while maintaining the PESQ at the same level with DCCRN. 2) We update the network structure with convolution based connections between encoder and decoder layers for better embedding and complex TF-LSTM for temporal dependency modeling. This update brings an extra PESQ gain of 0.05. 3) To maintain good speech quality, we formulate the decoder as a multi-task learning framework with an auxiliary task of \textit{a priori} SNR estimation which has proven to be beneficial to the perceived speech quality~\cite{nicolson2020masked}. This benefit is confirmed with PESQ gain of 0.03 according to our experiments. 4) Finally the new DCCRN+ model (with MMSE-LSA based post-processing for residual noise removal) has surpassed several state-of-the-art models and obtained superior performance with 3.51 MOS in the DNS2021 challenge.

    \begin{figure*}[t]
    \centering
    \includegraphics[width=1.0\linewidth]{ 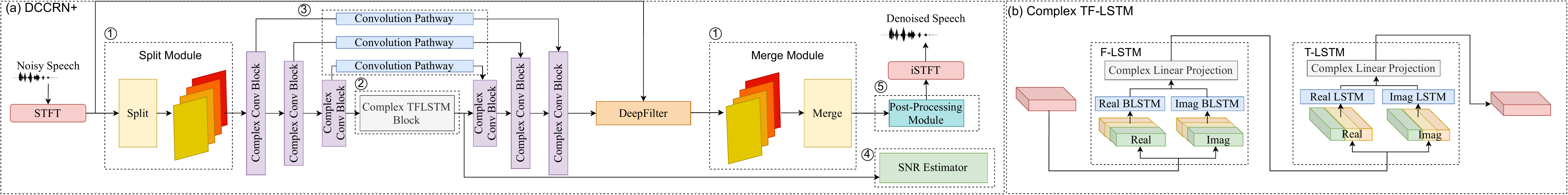}
    \vspace{-0.5cm}    
    \caption{Illustration of (a) Network structure of the proposed DCCRN+ and (b) Complex TF-LSTM Block. Main contributions include 1) subband processing, 2)  complex TF-LSTM, 3) convolution pathway, 4)  SNR estimation module and 5) post-processing.}
    \label{fig:total}
    \vspace{-0.6cm}    
    \end{figure*}

\section{DCCRN+}
\subsection{The new design}

DCCRN~\cite{hu2020dccrn} is the complex network version of CRN with a symmetrically-designed encoder-decoder structure and long short-term memory (LSTM) layers. The encoder and decoder are composed of complex convolution/deconvolution layers and LSTM or complex LSTM is inserted in between, which intends to model temporal dependencies in speech. The complex module models the correlation between magnitude and phase with the simulation of complex multiplication. Skip connections are used to concatenate the output of each encoder layer to the input of the corresponding decoder layer. With this model structure, DCCRN has achieved the best MOS in the subjective listening test in Interspeech 2020 deep noise suppression challenge.

We substantially extend DCCRN with several important revisions. The model structure of the so-called DCCRN+ is shown in Figure~\ref{fig:total}(a). We can see that the general structure is similar to DCCRN but with the following differences:
1) Subband processing with split/merge modules before/after encoder/decoder. 2) Complex TF-LSTM for both frequency and time scale temporal modeling. 3) Addition of convolution for better information aggregation from encoder output before feeding to decoder. 4) Addition of SNR estimation module to alleviate speech distortion during noise suppression. 5) Post-processing to further remove residual noise.

    \subsection{Subband Processing}
    
    
    Subband processing is a common approach in audio processing~\cite{li2020online, liu2020channel, yang2021multi} to reduce model size and save computation cost.
    The audio in each frequency subband can be down-sampled by a factor of $K$ (the number of frequency bands), and thus the total computational cost can be reduced. Previous work~\cite{takahashi2017multi} also pointed out that local patterns in the spectrogram are often different in each frequency band: the lower frequency band tends to contain high energies, tonalities as well as long sustained sounds, while the higher frequency band is likely to have low energy components, noise and rapidly decaying sounds. Subbands are often split by engineered Finite Impulse Response (FIR) filters~\cite{yu2019durian} and it is difficult to design perfect filters for different applications. Recent attempts on learnable front-end using neural networks~\cite{zeghidour2021leaf} have shown promising results, which may avoid the troublesome manual design of specific filters. This inspires us to develop learnable subband split and merge modules automatically using neural layers.

    
    
    The input of the band-splitting module is the T-F spectrum resulted from short time Fourier transform (STFT), denoted as ${Y \in R^{T\times F}}$, where $T$ is the number of frames and $F$ is the number of frequency bins. We employ neural network based analysis filters ${A_{k}(f_{k})}$ for band splitting, where ${k\in 1, \cdots ,K}$ stands for the number of subbands and ${f_{k}\in (F/K)\cdot (k-1),\cdots ,(F/K)\cdot k}$. The output after split analysis for band $k$ is
    {
    \vspace{-0.3cm}
    \setlength\abovedisplayskip{1pt}
    \setlength\belowdisplayskip{1pt}
     \begin{equation}\label{split_band}
    \vspace{-0.1cm}
    \begin{aligned}
    Y_{k}(f, t) = \sum_{f_k=(F/K) \cdot (k-1)}^{(F/K) \cdot k}Y(f,t)\cdot A_{k}(f_{k}).
    \end{aligned}
    \vspace{0.05cm}
    \end{equation} 
    }Then ${Y_{k}}$ is processed by Instance Normalization to accelerate the model convergence. and maintain the independence between each band. ${Y_{k}}$ for band $k$ will go through the encoder-decoder structure, as shown in Fig. \ref{fig:total}(a). Deepfilter~\cite{mack2019deep} is adopted to process estimated mask and input subbands. 
    Finally they are merged back using
    {
    \setlength\abovedisplayskip{1pt}
    \setlength\belowdisplayskip{1pt}
    \begin{equation}\label{merge_band}
    \vspace{-0.1cm}
    \begin{aligned}
    \hat{X}(f,t)=\sum _{f}\text{CAT}(\hat{X}_{1},\cdots ,\hat{X}_{k},\cdots,\hat{X}_{K})\cdot S(f)
    \end{aligned}
    \vspace{-0.1cm}
    \end{equation}      
    }where ${\hat{X}_{k}=\hat{X}_{k}(f_{k}, t)}$ denotes the network output of ${k}$ subband. After concatenating them together, we use another neural network based synthesis filters ${S(f)}$ to obtain the (merged) output spectrum ${\hat{X}(f, t)}$. The number of bands $K$ can be pratically set, and ${A_{k}(f_{k})}$ and ${S(f)}$ are learnable neural network layers.
    \vspace{-0.1cm}
    \subsection{Complex TF-LSTM Block}
    DCCRN employs LSTM layers to process the encoder output for temporal dependency modeling at time scale. Inspired by~\cite{li2015lstm}, we explicitly model the frequency-wise evolution of spectral patterns as well by another LSTM module Fig.~\ref{fig:total}(b). Specifically, we first use a complex frequency-LSTM (F-LSTM) to browse the frequency bands so that frequency-evolving information is summarized, and then the output layer activations are used as the input to the complex time-LSTM (T-LSTM) for time scale summarization.
    In details, the complex F-LSTM can be described as
    {
    \setlength\abovedisplayskip{1pt}
    \setlength\belowdisplayskip{0.5pt}
    \begin{equation}\label{complex_TF-LSTM_f_lstm}
    \begin{aligned}
    U_{f} = [\text{CAT}(\text{BLSTM}_{r}(\Re{(E)}[i,:,:]),\\\text{BLSTM}_{}(\Im{(E)}[i,:,:])), i=1\cdots F]
    \end{aligned}
    \end{equation}      
    }where ${E \in R^{F\times T \times C}}$ denotes the encoder output. Then we send the real($\Re{}$)/imag($\Im{}$) parts of ${E}$ to real/imag BLSTMs separately. After concatenating (CAT in Eq~(\ref{complex_TF-LSTM_f_lstm})) them together, the output $U_f$ is processed by Complex Linear Projection (CLP)~\cite{variani2016complex}) without modulo operation: 
    \setlength\abovedisplayskip{1pt}
    \setlength\belowdisplayskip{1pt}
    \begin{equation}\label{complex_TF-LSTM_f_clp}
    \begin{aligned}
    O_{f} = \text{CLP}(U_{f}).
    \end{aligned}
    \end{equation}  
    Subsequently we take the outputs from the complex F-LSTM as the input to feed to the complex T-LSTM to do time scale analysis:
    {
    \setlength\abovedisplayskip{1pt}
    \setlength\belowdisplayskip{1pt}
    \begin{equation}\label{complex_TF-LSTM_t}
    \begin{aligned}
    O_{t} = \text{CLP}([\text{CAT}(\text{LSTM}_{r}(\Re{(O_{f})}[:,i,:]),\\\text{LSTM}_{i}(\Im{(O_{f})}[:,i,:])), i=1\cdots T])
    \end{aligned}
    \end{equation} 
    }where $O_{t}$ is the final output of the complex TF-LSTM block which is subsequently sent to the decoder. Note that we use unidirectional LSTM at time scale as real-time speech enhancement is time sensitive.
    
    \vspace{-0.2cm}
    
    \subsection{Convolution Pathway}
    In the original DCCRN, through skip-connections, the feature maps of the encoder are directly concatenated to the decoder. In DCCRN+, before feeding to the decoder layers, the encoder output in each layer undergoes a convolution-based information aggregation block. Specifically, the skip pathway between encoder and decoder consists of a complex convolution block and batch normalization. 
    
    \subsection{SNR Estimator}
    Our aim is to maintain good speech quality while reducing noise. Previous experiences~\cite{zheng2020interactive} show that directly training the neural noise suppression module may inevitably lead to a certain amount of speech distortion. We migrate this problem by using an SNR estimator to estimate frame-level SNR under a multi-task learning framework. The input of SNR estimator is Complex TF-LSTM, then it is fed to one LSTM layer and a Conv1D layer with sigmoid to estimate frame-level SNR. This module is used as an auxiliary task in training and removed at inference stage. 
    The SNR target label is given as follows:
    {
    \setlength\abovedisplayskip{1pt}
    \setlength\belowdisplayskip{1pt}
    \begin{equation}\label{snr_label_db}
    \begin{aligned}
    \xi(t)  = 20log_{10}(X(t) / N(t))
    \end{aligned}
    \end{equation}
    }where ${t}$ is the frame index, ${\xi(t)}$ is the ${priori}$ frame-level SNR, ${X(t)}$ is the clean speech spectrum and ${N(t)}$ is the noise spectrum. To avoid the influence of the fluctuation of SNR estimation, a simple processing is done on the ${priori}$ frame-level SNR during training.
    {
    \setlength\abovedisplayskip{1pt}
    \setlength\belowdisplayskip{1pt}
    \begin{equation}\label{snr_label_first}
    \vspace{-0.1cm}
    \begin{aligned}
    \hat{\mu} &= \hat{\mu}\cdot  \alpha + \mu \cdot(1-\alpha ) \\
    \hat{\sigma } &= \hat{\sigma }\cdot  \alpha + \sigma \cdot(1-\alpha )
    \end{aligned}
    \end{equation}    
    }where ${\hat{\mu}}$ and ${\hat{\sigma}}$ denote the moving average of the SNR mean and std respectively. The current SNR mean and std of a training utterance is denoted as ${\mu, \sigma}$. Then the SNR is normalized.
    {
    \setlength\abovedisplayskip{0pt}
    \setlength\belowdisplayskip{0pt}
    \begin{equation}\label{snr_label_second}
    \vspace{-0.01cm}
    \begin{aligned}
    \xi(t) &= (\xi{(t)} - \hat{\mu})/\hat{\sigma}
    \end{aligned}
    \vspace{-0.01cm}
    \end{equation}  
    }
    After that, we use ERF (Error Function) and compress SNR to 0 $\sim$ 1 to avoid the too large range which may lead to poor network convergence~\cite{nicolson2020masked}. In this paper, we empirically set ${\alpha}$ to 0.99. 
    {
    \setlength\abovedisplayskip{1pt}
    \setlength\belowdisplayskip{1pt}
    \vspace{-0.4cm}
    \begin{equation}\label{snr_label_second}
    \begin{aligned}
    \xi(t) &= (\text{ERF}(\xi(t)) + 1)/2
    \end{aligned}
    \vspace{-0.95cm}
    \end{equation}
    }

\vspace{0.4cm}
\subsection{Post-Processing}
    With the SNR estimator, despite the notable improvement of speech quality, there may exist some residual noise. In order to remove such residual noise, we adopt a post-processing module.
    The estimated spectrum and noisy spectrum are used to calculate ${priori}$ and ${posterior}$ SNR as follows:
    {
    \setlength\abovedisplayskip{1pt}
    \setlength\belowdisplayskip{1pt}
    \begin{equation}\label{snr_label_second}
    \begin{aligned}
    \xi'(t) &=\text{CDF}(\sigma^{2}_{\hat{x}}(t))/\text{CDF}(\sigma^{2}_{n}(t)) \\
    \gamma(t) &=\text{CDF}(\sigma^{2}_{y}(t))/\text{CDF}(\sigma^{2}_{n}(t))
    \end{aligned}
    \end{equation} 
    }where ${\xi'(t), \gamma(t)}$ are the ${priori}$ and ${posterior}$ frame-level SNR, ${\text{CDF}}$ is the cumulative distribution function, ${\sigma^{2}_{\hat{X}}(t)}$ is the variance of estimated spectrum, and $\sigma^{2}_{N}(t)$ and ${\sigma^{2}_{Y}(t)}$ are the variance of noise spectrum and noisy spectrum respectively. With the estimated SNR, an MMSE-LSA estimator~\cite{ephraim1985speech} is introduced to compute the final frame-level gain ${\text{G}}$ and then applied to estimated spectrum ${\hat{X}}$ to suppress the residual noise.
    \vspace{-0.3cm}
    {
    \setlength\abovedisplayskip{1pt}
    \setlength\belowdisplayskip{1pt}
    \begin{equation}\label{post_process}
    \vspace{-0.2cm}
    \begin{aligned}
    \text{G} &= \text{MMSE\_LSA}(\xi', \gamma) \\
    \hat{X} &= \text{G} \cdot \hat{X}
    \end{aligned}
    \end{equation}     
    }In practice, the noise suppression ability of the post-processing module gradually deteriorates because of cumulative variance. Thus we use the rate-of-change of SNR to reset the variance:
    {
    \setlength\abovedisplayskip{1pt}
    \setlength\belowdisplayskip{1pt}
    \begin{equation}\label{post_process}
    \begin{aligned}
    r(t) &= (\xi'(t)-\xi'(t-1))/\xi'(t-1)
    \end{aligned}
    \end{equation}
    }where ${r(t)}$ is the rate-of-change of SNR. Once ${r(t)}$ is larger than 1, the variance is reset. In this way, we can observe that the speech is undamaged and residual noise is largely removed.
    
    \subsection{Loss Function}
    The SI-SNR loss~\cite{luo2019conv} is used in noise suppression. In addition, we also use MSE loss to guide the learning of SNR estimator. 
    \vspace{-0.3cm}
    {
    \setlength\abovedisplayskip{1pt}
    \setlength\belowdisplayskip{1pt}
    \begin{equation}\label{snr loss}
    \begin{aligned}
    L_{\text{SNR}}=\text{MSE}(\hat{\xi}, \xi)
    \end{aligned}
    \end{equation}  
    }where ${\hat{\xi}}$ is the output of the SNR estimation module and ${\xi}$ is the label. The final loss is 
    {
    \setlength\abovedisplayskip{1pt}
    \setlength\belowdisplayskip{1pt}
    \begin{equation}\label{loss}
    \begin{aligned}
    L = L_{\text{SI-SNR}} + \delta \cdot L_{\text{SNR}}.
    \end{aligned}
    \end{equation}  
    }In this paper, as the values of the two losses are not in the same scale, we empirically set ${\delta}$ to 30.
    
\section{Experiments}
    
    \subsection{Datasets}
    We first take a comprehensive ablation study on the proposed model on the DNS-2020 dataset~\cite{reddy2020interspeech}. Then our model is trained, integrated with the post-processing module and evaluated with the Interspeech 2021 DNS challenge dataset (DNS-2021)~\cite{reddy2021interspeech} to show its performance on more complicated and real acoustic scenarios. We also compare other competitive models (such as PercepNet~\cite{valin2020perceptually}) with our model on Voice Bank + DEMAND dataset~\cite{valentini2016investigating} as these models have PESQ scores reported on this set.
    Specifically, DNS-2020 speech set includes a total of 500 hours of clean speech from 2,150 speakers. The 180-hour DNS-2020 noise set includes 65,000 noise clips from 150 noise classes. And there are 80,000 RIR clips, coming from the RIR\_NOISES set~\cite{ko2017study} and the provided RIR set by the challenge. For DNS-2021 dataset, we drop the utterances with poor quality. Totally, we generate a large clean dataset with 760 hours duration. The training set and validation set are configured according to a ratio of 9:1 after shuffling. In the training stage, we perform dynamic mixing of speech and noise, while the SNR ranges from -5db to 20db, and there is 50\% probability of convolving with RIR for simulating speech with reverberation. We also apply a biquad filter~\cite{valin2018hybrid} on half of the training data. As for the Voice Bank + DEMAND dataset, there are 824 samples from 8 speakers for testing.
    
    
    \subsection{Training setup and baselines}
    For all of our models, the window length and frame shift are 20ms and 10ms respectively and the FFT length is 512. We use Adam optimizer and the initial learning rate is 1e-3. When the loss of the validation set increases, the learning rate will be decayed by a ratio of 0.5. 
   
        \textbf{DCCRN}: The number of channels for the DCCRN is \{16,32,64,128,256,256\}, and the convolution kernel and step size are set to (5,2) and (2,1). Two LSTM layers are adopted and the number of nodes is 256. There is a 1024*256 fully connected layer after the LSTM. Each encoder module handles the current frame and one previous frame. In the decoder, the last layer processes one extra future frame, and each previous layer utilizes the current frame and one historical frame.

        \textbf{DCCRN+}: The number of channels for DCCRN+ is \{32,64,128,256\}. The split-band module is a group Conv1D layer with 4 groups. Correspondingly, the merge-module is a linear layer. The Complex TF-LSTM module is composed of one complex LSTM (the units of real and imag parts are 256 separately) and one complex BLSTM (real and imag parts are 256 for each direction separately). The CLP module has 256 units for real and imag parts. The convolution pathway module is composed of a $1\times1$ complex Conv2D layer. The SNR Estimator is a 64 units LSTM layer followed by a Conv1D layer with 3 kernels. The rest configuration is the same as DCCRN.

    Since only 10ms future speech is seen in the last layer of the decoder, the total processing time is ${20+10+10=40ms}$, which meets the challenge requirement: frame length + frame shift + future frame ${\leq}$ 40ms.
    \subsection{Experimental results and discussion}
    PESQ~\cite{rec2005p} is first reported for various models on the DNS-2020 synthetic test set, as shown in Table~\ref{tab:dns_2020pesq}. Real Time Factor (RTF) is also tested on a machine with an Intel(R) Xeon(R) CPU E5-2640@2.50GHz in single thread. Note that NSNet is the baseline model provided by the challenge organizer.

    \begin{table}[t]
    \footnotesize
    \setlength\tabcolsep{1pt}
    \caption{Various models' PESQ on DNS-2020 synthetic test set.}
    \vspace{-0.2cm}
    \begin{tabular}{lccccccc}
    \toprule 
    Model   & Para.(M) & \tabincell{c}{look-\\ahead\\(ms)}&  RTF & \tabincell{c}{no\\reverb}& reverb  & Ave. \\ 
    \midrule 
    Noisy           & -  & 0  & -& 2.45     & 2.75   & 2.60\\ 
    NSNet~\cite{xia2020weighted} (Baseline) &  1.3 & 10 & -& 3.07     & 2.81   & 2.94\\
    DCCRN    &  3.7  & 10 & 0.272 & 3.26   & 3.20   & 3.23\\ 
    Subband DCCRN (FIR filter) &  2.8  & 10 & 0.177 & 3.03     &  3.11  & 3.07\\
    Subband DCCRN (NN filter) &  2.8  & 10 & \textbf{0.137} & 3.25     &  3.22  & 3.24\\ 
    +Complex TF-LSTM &  3.3 & 10 &0.244 &  3.29   & 3.27   &  3.28    \\
    ~ + Convolution Pathway&  3.3 & 10 & 0.248 & 3.30     &  3.28  & 3.29\\
    ~ ~ + SNR Estimator &  3.3 & 10 & 0.248 & \textbf{3.33}     &  \textbf{3.30}  & \textbf{3.32} \\
    ~ ~ ~ + Post-processing &  3.3 & 10 & 0.250 &  3.27  & 3.23  & 3.25 \\
    \bottomrule
    \label{tab:dns_2020pesq}
    \end{tabular}
    \vspace{-0.5cm}
    \end{table}
    
\begin{figure}[t]
    \centering
    \includegraphics[width=.9\linewidth]{ 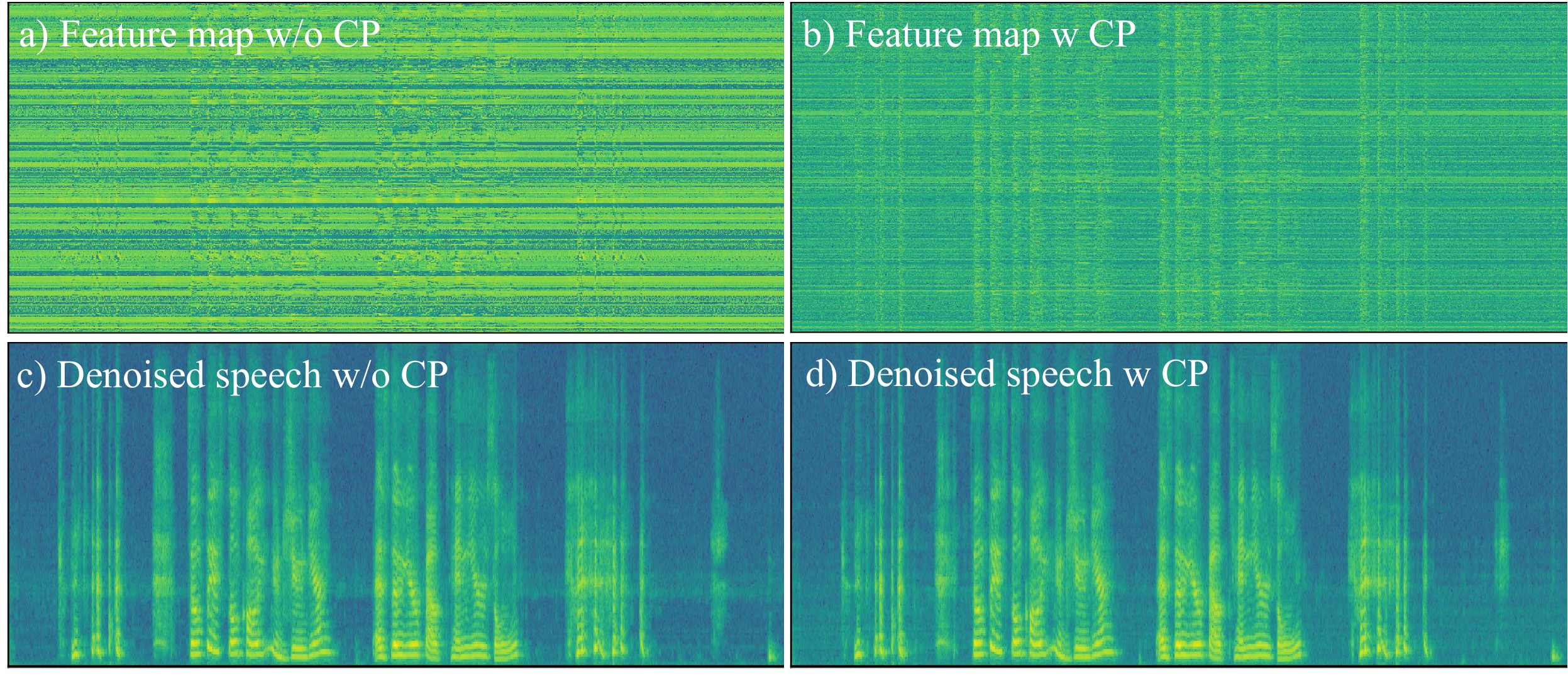}
    \vspace{-0.3cm}    
    \caption{Comparison on the denoising result on a testing noisy clip for the cases with and without convolution pathway (CP).}
    \label{fig:compare_of_sc_module}
    \end{figure}

    \begin{table}[t]
    \centering
    \setlength\tabcolsep{2pt}
    \caption{PESQ on Voice Bank + DEMAND}
    \vspace{-0.2cm}
    \begin{tabular}{lcccc}
    \toprule 
    Model   & Para.(M) & \tabincell{c}{Semi-\\Causal} & \tabincell{c}{External\\Data }&PESQ-WB \\ 
    \midrule 
    Noisy       & - & - & - &1.97 \\ 
    RNNoise~\cite{valin2018hybrid} & 0.06 & $\surd$ & $\surd$ &2.29 \\ 
    PercepNet~\cite{valin2020perceptually}  & 8 & $\surd$ &  $\surd$ &2.73 \\ 
    DCCRN & 3.7 & $\surd$ & $\surd$ &2.68 \\
    DCCRN+ & 3.3 & $\surd$ & $\surd$ &\textbf{2.84} \\
    \bottomrule
    \label{tab:VoiceBank_DEMAND_pesq}
    \end{tabular}
    \vspace{-1cm}
    \end{table}
    The result in Table~\ref{tab:dns_2020pesq} shows that the subband operation can considerably improve speed and reduce model size. But the PESQ for the FIR-filter-based subband DCCRN model has a clear degradation with the orginal DCCRN. With the help of the proposed neural network filter, the PESQ for the subband model is restored to the same level with DCCRN and the inference speed is further boosted with RTF of 0.137.
    After substituting LSTM with complex TF-LSTM in subband DCCRN (NN filter), we obtain a noticeable PESQ improvement, while the model becomes larger and slower. The use of convolution pathway and SNR estimator brings further PESQ gain and the best PESQ is 3.32, which is a clear improvement as compared with original DCCRN. Interestingly, when comparing the feature map before and after using the convolution pathway (CP), we find that the use of CP results 
   in more clear patterns on the feature map, leading to better noise reduction. A case study on one testing clip is shown in Fig.\ref{fig:compare_of_sc_module}, where more noise is suppressed when CP is applied. Finally, when post-processing is applied, PESQ is decreased due to some spectrum information lost~\cite{li2021icassp}. However the use of post-processing is beneficial to subjective listening as shown in previous works~\cite{valin2020perceptually,li2021icassp} because unnatural residual noise is further suppressed.

    We compare the PESQ performance of DCCRN+ with other competitive models on the Voice Bank + DEMAND test set. The results in Table~\ref{tab:VoiceBank_DEMAND_pesq} show that the proposed DCCRN+ clearly outperforms other models, DCCRN+ surpasses PercepNet by a large margin with even fewer parameters. 
    
    We further test our models trained on the DNS-2021 dataset. This time we use DNSMOS~\cite{reddy2021dnsmos} for evaluation -- the new metric provided by the challenge organizer that is believed to be more correlated with subjective listening score. As can be seen from Table~\ref{tab:dnsmos} that DNSMOS is increased with the addition of the updates on our model and the highest score 3.46 is achieved by the use of all updates, including the post-processor. We submitted the enhanced clips on the blind test set using the the final model. Table~\ref{tab:dns_mos} shows the P.835~\cite{naderi2020crowdsourcing} DNS-2021 subjective evaluation results released by the challenge organizer. We can observe that our DCCRN+ model significantly outperforms NSNet2 baseline with large MOS improvement. Our system ranked top 4 in overall MOS in all 20 submissions in the wide-band track in the challenge. 
    
    
    \vspace{-0.5cm}
    \begin{table}[t]
    \centering
    \footnotesize
    \setlength\tabcolsep{10pt}
    \caption{DNSMOS on DNS-2021 blind test set.}
    \vspace{-0.3cm}
    \begin{tabular}{lcc}
    \toprule 
    Model   & DNSMOS \\ 
    \midrule 
    Noisy       & 2.94 \\ 
    DCCRN  &  3.40 \\ 
    Subband DCCRN (FIR filter) & 3.16 \\
    Subband DCCRN (NN filter) & 3.37 \\
    ~ + complex TF-LSTM & 3.40\\
    ~~ + CP & 3.42 \\
    ~~~ + SNR Estimator & 3.43 \\
    ~~~~ + post-processing & \textbf{3.46} \\
    \bottomrule
    \label{tab:dnsmos}
    \end{tabular}
    \vspace{-0.7cm}
    \end{table}
    
    \vspace{0.2cm}
    \begin{table}[htbp]
    \footnotesize
    \setlength\tabcolsep{0.5pt}
    \caption{DMOS on DNS-2021 challenge blind test set}
    \vspace{-0.3cm}
    \begin{tabular}{lcccccccc}
    \toprule 
    Model & Stat. & Emot. & Non-En-T & Non-En & Mus & En. & Overall \\
    \midrule 
    Noisy & 0(3.03) & 0(2.28) & 0(3) & 0(3.04) & 0(2.57) & 0(2.52) & 0(2.77) \\ 
    NSNet2~\cite{xia2020weighted} & 0.25 & 0.47 & 0.31 & 0.21 & 0.21 & 0.41 & 0.30 \\
    DCCRN+ &  \textbf{0.69} & \textbf{1.13} & \textbf{0.59} & \textbf{0.64} & \textbf{0.76} & \textbf{0.75} & \textbf{0.74} \\
    \bottomrule
    \label{tab:dns_mos}
    \end{tabular}
    \vspace{-0.8cm}
    \end{table}    

\section{Conclusions}
\label{sec:typestyle}
    This paper presents substantial updates on our previously proposed DCCRN model for speech enhancement. The new model, called DCCRN+, is equipped with subband processing ability via learnable neural filters for band split and merge, leading to compact model size and speed-up inference. The new model has also been updated with TF-LSTM and convolution pathway. Importantly, an SNR estimator is adopted along with the decoder, under the multi-task learning framework, to maintain good speech quality while removing noise. Finally a post-processor is adopted to remove unnatural residual noise. Experiments have shown the effectiveness of these updates. The DCCRN+ model has achieved superior performance in the subjective listening test in the Interspeech2021 DNS challenge. Some of the enhanced audio clips on the blind test set can be found from \url{https://imybo.github.io/dccrn-plus/}.

    
\bibliographystyle{IEEEtran}

\bibliography{mybib}


\end{document}